\input epsf
\input phyzzx
\hsize=40pc



\catcode`\@=11 
\def\NEWrefmark#1{\step@ver{{\;#1}}}
\catcode`\@=12 

\def\square{\kern1pt\vbox{\hrule height 1.2pt\hbox{\vrule width 1.2pt\hskip 3pt
   \vbox{\vskip 6pt}\hskip 3pt\vrule width 0.6pt}\hrule height 0.6pt}\kern1pt}
\def\inbar{\,\vrule height 1.5ex width .4pt depth0pt}\def\IC{\relax
\hbox{$\inbar\kern-.3em{\rm C}$}}
\def\IH{\relax\hbox{$\inbar\kern-.3em{\rm H}$}}

\def\bra#1{\langle #1 |}
\def\ket#1{| #1 \rangle}

\def\A{{\cal A}}
\def\B{{\cal B}}

\def\I{{\cal I}}

\def\K{{\cal K}}

\def\O{{\cal O}}

\def\V{{\cal V}}

\def\p{\partial}

\def\wt{\widetilde}
\def\wh{\widehat}

\def\B{{\cal B}}

\def\V{{\cal V}}
\def\O{{\cal O}}

\def\p{\partial}

\singlespace

\def\define#1#2\par{\def#1{\Ref#1{#2}\edef#1{\noexpand\refmark{#1}}}}
\def\con#1#2\noc{\let\?=\Ref\let\<=\refmark\let\Ref=\REFS
         \let\refmark=\undefined#1\let\Ref=\REFSCON#2
         \let\Ref=\?\let\refmark=\<\refsend}

\let\refmark=\NEWrefmark

\define\csft{B. Zwiebach, `Closed string field theory: Quantum action and the
Batalin-Vilkovisky master equation',
Nucl. Phys. {\bf B390} (1993) 33, hep-th/9206084.}

\define\manifest{S. Rahman, `The path towards manifest background
independence', MIT-CTP-2649, {\it to appear.}}

\define\recur{S. Rahman, `String vertices and inner derivations',
MIT-CTP-2645, June 1997, hep-th/9706128.}

\define\cbi{A. Sen and B. Zwiebach, `Local background independence of
classical closed string field theory',
Nucl. Phys. {\bf B414} (1994) 649, hep-th/9307088.}

\define\qbi{A. Sen and B. Zwiebach, `Quantum background independence of closed
string field theory',
Nucl. Phys. {\bf B423} (1994) 580, hep-th/9311009.}

\define\nonconf{B. Zwiebach, `Building string field theory around
non-conformal backgrounds',
Nucl. Phys. {\bf B480} (1996) 541, hep-th/9606153.}

\define\dilaton{O. Bergman and B. Zwiebach, `The dilaton theorem and closed
string backgrounds',
Nucl. Phys. {\bf B441} (1995) 76, hep-th/9411047.}

\define\sabdilaton{S. Rahman and B. Zwiebach, `Vacuum vertices and the
ghost-dilaton',
Nucl. Phys. {\bf B471} (1996) 233, hep-th/9507038.}

\define\basisone{S. Rahman, `Consistency of quantum background independence',
MIT-CTP-2626, April 1997, hep-th/9704141.}

\singlespace
{}~ \hfill \vbox{\hbox{MIT-CTP-2648}
\hbox{
} }\break
\title{GEOMETRISING THE CLOSED STRING FIELD THEORY ACTION}
\author{Sabbir A Rahman \foot{E-mail address: rahman@marie.mit.edu
\hfill\break Supported in part by D.O.E.
cooperative agreement DE-FC02-94ER40818.}}
\address{Center for Theoretical Physics,\break
Laboratory of Nuclear Science\break
and Department of Physics\break
Massachusetts Institute of Technology\break
Cambridge, Massachusetts 02139, U.S.A.}

\abstract
{We complete the set of string vertices of non-negative dimension by
introducing in a consistent manner those moduli spaces which had previously
been excluded. As a consequence we obtain a `geometrised' string action taking
the simple form $S = f(\B)$ where `$\B$' is the sum of the string vertices.
That the action satisfies the B-V master equation follows from the recursion
relations for the string vertices which take the form of a `geometrical'
quantum master equation.}

\endpage
\singlespace
\baselineskip=18pt

\chapter{\bf Introduction and Summary}

One hopes eventually to be able to reformulate closed string field theory
in a form which is both simplified and which brings out the deeply geometrical
underlying basis. There were two main hurdles which needed to be overcome
before such a goal could be realised. The first of these was the need to find
a geometrical description of the usual string field theory operators $\p$,
$\K$ and $\I$. The second was the need to complete the set of string vertices
$\B^{\bar n}_{g,n}$ by both introducing those vertices of non-negative
dimension which had for various reasons previously been excluded, and extending
the set to include the vertices of `negative' dimension\foot{Recall that the
dimension of $\B^{\bar n}_{g,n}$ is $6g-6+2n+3\bar n$.}.

In an earlier paper [\recur], some advance was made in this
direction. It was shown there that it was consistent to express all three
operators, $\p$, $\K$ and $\I$ as inner derivations on the  B-V algebra of
string vertices. Indeed the following identifications were derived,
$$\eqalign{\p \A &= \{ L , \A \}\cr
\K \A &= - \{ \A , R \}\cr
\I \A &= \{ \A , l \}\,,}\eqn\sabone$$
where $L$, $R$ and $l$ are graded-even elements of the B-V algebra satisfying,
$$\eqalign{L + R &= \B^0_{0,2}\cr
l &= \B^1_{0,1}\,,}\eqn\sabtwo$$
the moduli space $\B^1_{0,1}$ representing a sphere with one ordinary and
one special puncture, and $\B^0_{0,2}$ a sphere with two ordinary
punctures. The spaces $L$, $R$ and $l$ were additionally required to satisfy
the following constraints,
$$\{ L , L \} = \{ L , l \} = \{ l , l \} = \{ R , R \} = 0\,.\eqn\sabthree$$
The recursion relations for the string vertices were then given by a
`geometrical' quantum B-V master equation,
$$\half \{ \B , \B \} + \Delta \B = 0\,,\eqn\sabfour$$
while the action took the form,
$$S = S_{1,0} + f(\B)\,.\eqn\sabfive$$
While this signalled firm progress, it was not yet totally satisfactory seeing
as $\B = \sum_{g,n,\bar n} \B^{\bar n}_{g,n}$ still excluded the
non-negative dimensional vertices $\B^1_{1,0}$, $\B^0_{1,0}$ and
$\B^{\bar n}_{0,0}$ (where $n\geq2$). The main goal of the current paper will
be to include these spaces in a consistent manner ensuring that previous
results (particularly quantum background independence [\qbi], the
ghost-dilaton theorem [\dilaton,\sabdilaton], and the recursion relations
[\csft]), are still satisfied.

\noindent
The structure of our paper is as follows.

In \S 2.1 we review the original reasons for excluding the spaces
$\B^0_{1,0}$, $\B^1_{1,0}$ and $\B^{\bar n}_{0,0}$, and explain why the
existence of the space $\B^0_{0,2}$ (which was introduced in [\recur]) makes
it consistent to reintroduce them. As a corllary we also see how a choice of
$\B^0_{0,2}$ determines the choice of connection. In \S 2.2 we briefly explain
why it is consistent to introduce the moduli spaces $\B^{\bar n}_{0,0}$ and
thereby complete the set of string vertices of
non-negative dimension.

Armed with the extended $\B$-complex we then state in \S 2.3 the expression
for the quantum action around arbitrary
string backgrounds, generalising the results of [\nonconf]. It takes a
completely geometrical form expressed solely as a function of the sum of
string vertices, $S = f(\B)$. Furthermore,
the recursion relations are contained in a quantum master action for the
$\B$-spaces. As an application of these we end by listing expressions for
the boundaries of the newly introduced moduli spaces.

In the conclusion in \S 4 we review what has been achieved. Unless explicitly
stated otherwise, we will use units in which $\hbar = \kappa = 1$ throughout.

\chapter{\bf String Vertices and the Geometrised Action}

In this section, we will extend the $\B$-space complex by introducing
the positive-dimensional vertices $\B^1_{1,0}$ (being a moduli
space of tori with a single special puncture), $\B^0_{1,0}$ (a space of
unpunctured tori), and $\B^{\bar n}_{0,0}$ where $n\geq2$ (a moduli spaces
of genus zero surfaces with no ordinary punctures and at least two special
punctures).

\section{The Moduli Spaces $\B^1_{1,0}$ and $\B^0_{1,0}$}

All of the spaces mentioned above are distinguished by the fact that they are
moduli spaces of surfaces containing no ordinary punctures. This means that
they cannot be sewn, and hence can take an active part neither in background
deformations (in particular they should not affect the proofs of background
independence or the ghost-dilaton theorem), nor in the recursion relations, so
one would expect that there
should be no problem in including them into the $\B$-space complex. Let us
then reconsider the reason for leaving out these spaces, and readmit them
if we find a satisfactory excuse for doing so. We first discuss briefly why the
space $\B^1_{1,0}$ was excluded.

When the spaces $\B^1_{g,n}$ were first introduced for the proof of quantum
background independence, they were defined only for $n\geq2$ at genus zero
and for $n\geq1$ at higher genus (Eqn.(4.19) of [\qbi]), the remaining spaces
$\B^1_{g,0}$ (for all genera $g\geq1$) being set to vanish as they could not
take part in background deformations and were therefore irrelevant to the
discussion.

However, their irrelevance to background independence was not a reason for
dismissing them entirely, and this was acknowledged in proving the
ghost-dilaton theorem [\dilaton,\sabdilaton], where all these spaces were
reinstated, with the
exception of the space $\B^1_{1,0}$. The reason for this omission was that
the general equation derived in [\dilaton] implied that its boundary
was given by
$\p \B^1_{1,0} = - \Delta \B^1_{0,2} - \I \B^0_{1,1}$,
which was inconsistent as it did not satisfy $\p \p \B^1_{1,0} = 0$.

This problem disappears if we assume the existence of the one loop vacuum
vertex $\B^0_{1,0}$ satisfying the recursion relations Eqn.\sabfour\ as we can
then show that $\p \p \B^1_{1,0} = 0$ is in fact satisfied. In particular,
we find the following,
$$\p \B^1_{1,0} = \K \B^0_{1,0} - \I \B^0_{1,1} - \Delta \B^1_{0,2}
\,.\eqn\sabfive$$
Taking the boundary once again and using the usual operator identities,
$$\eqalign{\p \p \B^1_{1,0} &= \p \K \B^0_{1,0} - \p \I \B^0_{1,1} - \p
\Delta \B^1_{0,2}\cr
&= \K \p \B^0_{1,0} - \I \p \B^0_{1,1} + \Delta \p \B^1_{0,2}\cr
&= - \K \Delta \B^0_{0,2} + \I \Delta \B^0_{0,3} + \Delta \K \B^0_{0,2}
- \Delta \I \B^0_{0,3}\cr
&= 0\,.}\eqn\sabsix$$
where we have applied the recursion relations to $\B^0_{1,0}$ and $\B^1_{0,2}$
to obtain the expressions $\p\B^0_{1,0}=-\Delta\B^0_{0,2}$ and $\p\B^1_{0,2}=
\K\B^0_{0,2}-\I\B^0_{0,3}$ for their boundaries. The term
$\K\B^0_{0,2}$ above should be identified with $\wt\V_{0,3}$ of \S 6.2 of
[\qbi], the usual $\V'_{0,3}$ being a special case. We shall discuss this in
more detail momentarily.

Having assumed the existence of $\B^0_{1,0}$ satisfying the recursion
relations, let us now
see further evidence to suggest why this is consistent. In particular, we
note that the proof of quantum background independence \S 6 of [\qbi] for the
field-independent $\O(\hbar)$ terms was particularly tricky, and left us with
the unusual condition $\p\B^0_{1,0}=-\pi_F \Delta\wt\V_{0,3}$ on the boundary
of $\B^0_{1,0}$. We will now show how our expectations for the one loop vacuum
vertex do away with these unwanted features.

\noindent
We recall the field-independent $\O(\hbar)$ condition for background
independence Eqn.(6.9) of [\qbi],
$$\p_\mu S_{1,0} = \int_{\I\B^0_{1,1}+\Delta\B^1_{0,2}} \bra{\Omega^{(0)1,1}}
\wh\O_\mu\rangle\,,\eqn\sabseven$$
Having assumed the existence of the vertex $\V_{1,0}$ we can now write this
as,
$$\p_\mu S_{1,0} = \p_\mu f(\V_{1,0}) = f_\mu(\K \V_{1,0}) = \int_{\K \V_{1,0}}
\bra{\Omega^{(0)1,1}} \wh\O_\mu\rangle\,.\eqn\sabeight$$
This allows us to rewrite the condition Eqn.\sabseven\ as follows,
$$\int_{\K\V_{1,0}-\I\V_{1,1}-\Delta\B^1_{0,2}} \bra{\Omega^{(0)1,1}}
\wh\O_\mu\rangle = 0\,.\eqn\sabnine$$
If we glance at Eqn.\sabfive\ we see that integration
region is simply $\p \B^1_{1,0}$. A simple application of Stokes' theorem
then explains why the background independence condition is satisfied,
$$\int_{\p\B^1_{1,0}} \bra{\Omega^{(0)1,1}} \wh\O_\mu\rangle =
\int_{\B^1_{1,0}} \bra{\Omega^{(0)1,1}} Q \ket{\wh\O_\mu} = 0\,.\eqn\sabten$$
There is no need for any further analysis or application of auxiliary
constraints.

This shows that it is algebraically consistent to introduce $\B^0_{1,0}$, and
we now give a geometrical description of what this algebra implies. It was
mentioned in [\csft] that $\B^0_{1,0}$ was `not constrained' by the recursion
relations. On assuming the existence of $\B^0_{0,2}$, which is a
twice-punctured sphere representing the kinetic term, this is no longer true
and, as we have mentioned, the recursion relations imply that $\p \B^0_{1,0} =
-\Delta \B^0_{0,2}$. This is as we would na\"ively expect - as we increase the
height of the internal foliation of the vacuum graph to $2\pi$, the diagram
should split into the twice-punctured sphere whose punctures are glued
together by a propagator [Fig. 1].

\midinsert
\epsfxsize 6in
\centerline{\epsffile{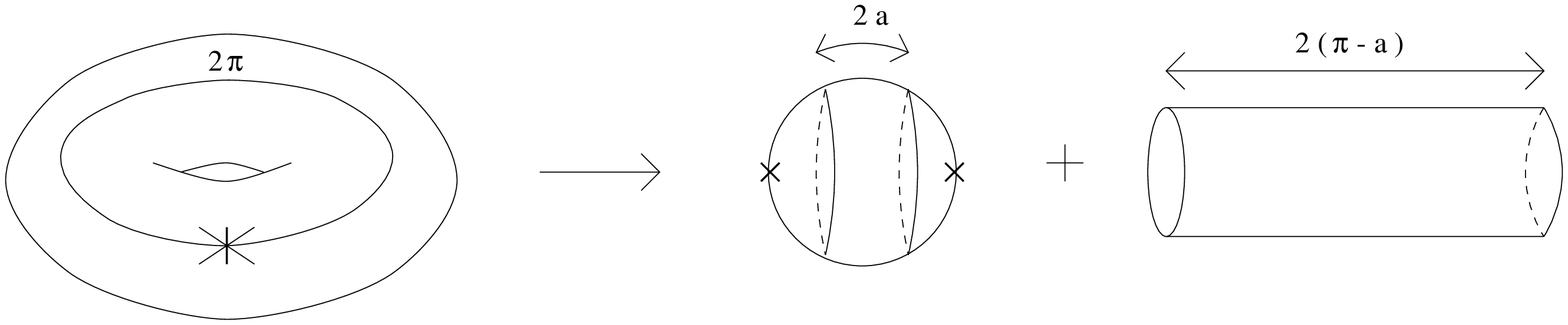}}
\centerline{Figure 1. The torus decomposition into a twice punctured sphere
and a propagator.}
\endinsert

Usually we consider the string vertices to have
stubs of length $\pi$, so that the propagator can have any length from $0$ to
$\infty$. 
In general it is also possible to use shorter stubs of length
$a$ say, where $0\leq a\leq\pi$, in combination with `cutoff propagators'
[\csft] which would be of length $\geq2(\pi-a)$. The latter condition would
ensure that no nontrivial loops have length less than $2\pi$, as is required
by the minimal area metric prescription. If we are to apply the same
conditions to the vertex $\B^0_{0,2}$ as are applied to the other vertices,
we would expect that it be conformally equivalent to a cylinder of length
$2a$. In this case, the general expression for the kinetic term would be,
$$Q = \half \bra{R_{12}} c_0^{-(2)} Q^{(2)} e^{2aL_0^{+(2)}} \ket{\Psi}_1
\ket{\Psi}_2\,.\eqn\sabeleven$$
For the usual case $a=\pi$, we require the insertion $e^{2\pi L_0^+}$.

\midinsert
\epsfxsize 5in
\centerline{\epsffile{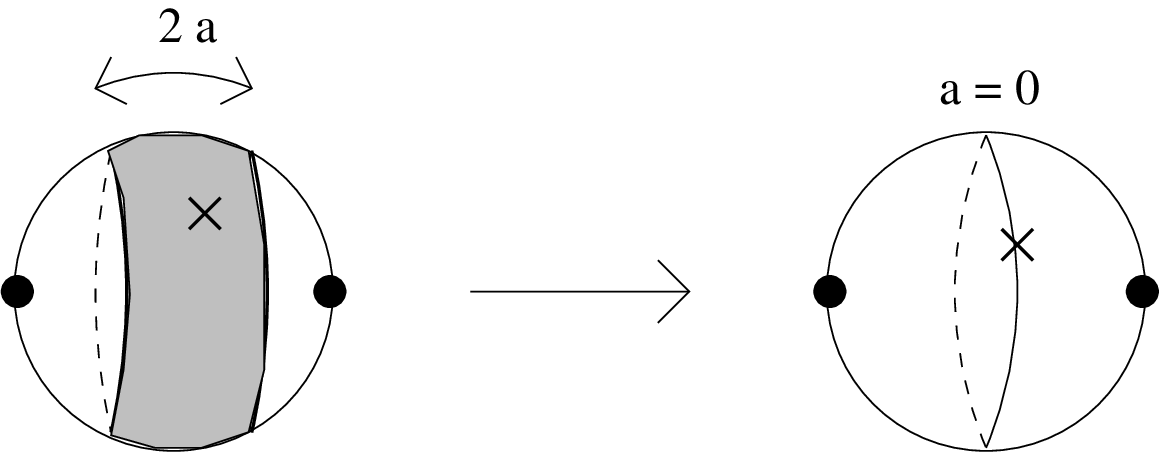}}
\centerline{Figure 2. The auxiliary vertex $\V'_{0,3}$ as a limiting case of
$\wt\V_{0,3}$.}
\endinsert

This description of $\B^0_{0,2}$ elucidates some points regarding the
choice of connection $\Gamma$. In \S 3.2 of [\basisone] we reviewed the origin
of symplectic connections. In particular, a choice of auxiliary three
string vertex $\wt\V_{0,3}$ determines the vertex $\B^1_{0,2}$ (which
interpolates between $\I \B^0_{0,3}$ and $\wt\V_{0,3}$), and this (or more
accurately, $\Delta \wt\V_{0,3}$), in turn
determines the choice of connection through Eqn.(3.23) of [\basisone]. The
particular case of the canonical connection $\wh\Gamma$ follows from choosing
$\V'_{0,3}$ (introduced in \S 3.3 of [\cbi]) as the auxiliary
three-string vertex.

Now the recursion relations determine that the boundary of $\B^1_{0,2}$ is
given by,
$$\p \B^1_{0,2} = \K \B^0_{0,2} - \I \B^0_{0,3}\,,\eqn\sabtwelve$$
which means that $\wt\V_{0,3}$ is identified with $\K \B^0_{0,2}$\foot{Note
that in [\recur], we identified $\K\B^0_{0,2}=-\{L,R\}$ with $\V'_{0,3}$. The
identification with $\wt\V_{0,3}$ is more general, and it is $\wt\V_{0,3}$
which should appear in the recursion relations Eqn.(2.37) of [\recur].}.
Moreover,
the choice of auxiliary three-string vertex determining the connection is
defined by the choice of $\B^0_{0,2}$ which as we saw above encodes the choice
of stub length and cutoff propagator. Now, $\B^0_{0,2}$ is just a cylinder
of length $2a$. The operator $\K$ adds a puncture over the entire surface of
this cylinder, resulting in the vertex $\wt\V_{0,3}$\foot{The connection is
well-defined as sewing together the two ordinary punctures results in
precisely the same once-punctured torus for each surface in $\wt\V_{0,3}$
irrespective of the position of the special puncture. This follows from the
usual translation symmetry of the torus.}. The vertex $\V'_{0,3}$
stems from the particular case $\lim_{a\to0}$ corresponding to a degenerate
cylinder (being the overlap surface corresponding to the standard
twice-punctured sphere), [Fig. 2].
In this case the special puncture may only be added over the boundary of the
coordinate disks. Each of the resulting three-punctured spheres are identical
(being related simply by a rotation), and we thereby recover the usual
description [\cbi] of $\V'_{0,3}$. It was recognised in \S 6.2 of [\qbi] that
the canonical connection $\wh\Gamma$ has some rather singular properties. We
are now able to identify the source of these properties with the degeneration
of the cylinder described by $\B^0_{0,2}$.

The considerations above suffice to show that it is consistent to include the
moduli spaces $\B^1_{1,0}$ and $\B^0_{1,0}$ into the string vertices.

\section{The Moduli Spaces $\B^{\bar n}_{0,0}$}

The moduli spaces of spheres with at least two special punctures and no
ordinary punctures were not considered solely because they contribute
ineffectual constants to the action, and were therefore not essential
to the usual discussions. Consequently we see no harm in including them and
they will be reinstated in what follows.

\section{The Geometrised String Action}

Having successfully introduced into $\B$ the missing string vertices of
non-negative dimension, we can now state the resulting expression for the
action. It takes the elegant `geometrised' form,
$$S = f(\B)\,,\eqn\sabseven$$
where $\B = \sum_{g,n,\bar n} \B^{\bar n}_{g,n}$, where $(g,n,\bar n)$
may take all values except when $g=0$ and $n+\bar n\leq1$. That this action
satisfies the master equation is clear in view of the form of the
recursion relations, which are still given by a quantum master equation for
the string vertices (Eqn.(2.36) of [\recur]),
$$\half \{ \B , \B \} + \Delta \B = 0\,,\eqn\sabeight$$
The boundaries of the newly introduced moduli spaces are summarised as follows,
$$\eqalign{\p \B^1_{1,0} &= \K \B^0_{1,0} - \I \B^0_{1,1} - \Delta \B^1_{0,2}
\,,\cr
\p \B^0_{1,0} &= - \Delta \B^0_{0,2}\,,\cr
\p \B^2_{0,0} &= 0\,,\cr
\p \B^{\bar n}_{0,0} &= \K \B^{\bar n-1}_{0,0} - \I \B^{\bar n-1}_{0,1}\,,
\qquad \qquad (\bar n>2)\,.}
\eqn\sabeleven$$

\chapter{\bf Conclusion}

We have seen how the moduli space $\B^0_{0,2}$ introduced in [\recur]
has provided the key to incorporating into the sum of string vertices $\B$ the
spaces $\B^1_{1,0}$ and $\B^0_{1,0}$ which, together with the spaces
$\B^{\bar n}_{0,0}$, complete the set of string vertices of non-negative
dimension.

The entire theory has been encoded elegantly in terms of the set of string
vertices, and the string fields $\ket{\Psi}$ and $\ket{F}$, which define the
function mapping the string vertices to the action. 

One might note that the three moduli spaces $\B^1_{0,0}$, $\B^0_{0,1}$ and
$\B^0_{0,0}$ remain to be defined. These will be
discussed in a forthcoming paper [\manifest] which will be concerned mainly
with new insights related to background independence stemming from our
geometrised formulation.

\ack
I would like to thank Barton Zwiebach for his useful comments,
and Chris Isham for allowing me to attend Imperial College,
London where some of this work was completed.

\refout

\bye